\documentclass{article}
\usepackage{graphicx}
\usepackage[round]{natbib}
\usepackage{xcolor}
\usepackage{hyperref}
\usepackage{authblk}
\usepackage[margin=3cm]{geometry}
\usepackage{pdfpages}
\usepackage{pdflscape}

\hypersetup{
    colorlinks,
    linkcolor={red!50!black},
    citecolor={blue!50!black},
    urlcolor={blue!80!black}
}

\newcommand{\var}[1]{\textbf{#1}}

\newcommand{\ywComment}[1]{}

\title{The BEAT-CF Causal Model: A model for guiding the design of trials and observational analyses of cystic fibrosis exacerbations}

\author[1,2]{\href{mailto:<steven.mascaro@monash.edu>?Subject=BEAT-CF+Causal+Model}{Steven Mascaro}{}}
\author[1,2]{Owen Woodberry}
\author[3]{Charlie McLeod}
\author[3]{Mitch Messer}
\author[4,5]{Hiran Selvadurai}
\author[6]{Yue Wu\thanks{Corresponding author: yue.wu1@sydney.edu.au}}
\author[3]{Andre Schultz}
\author[3,6]{Thomas L Snelling}

\affil[1]{%
    Bayesian Intelligence\\
    Upwey, Melbourne, Australia
}
\affil[2]{%
    Faculty of Information Technology\\
    Monash University\\
    Clayton, Melbourne, Australia
}
\affil[3]{
    The Kids Research Institute Australia\\
    Nedlands, Perth, Australia
}

\affil[4]{
    Department of Respiratory Medicine\\
    The Children’s Hospital at Westmead\\
    Sydney Children's Hospitals Network\\
    Westmead, Sydney, Australia
}

\affil[5]{
    Discipline of Paediatrics and Child Health\\
    University of Sydney\\
    Camperdown, Sydney, Australia
}

\affil[6]{%
    School of Public Health\\
    Faculty of Medicine and Health\\
    University of Sydney\\
    Camperdown, Sydney, Australia
}

\date{December 2025}

\begin{document}

\maketitle

\begin{abstract}
    Loss of lung function in cystic fibrosis (CF) occurs progressively, punctuated by acute pulmonary exacerbations (PEx) in which abrupt declines in lung function are not fully recovered. A key component of CF management over the past half century has been the treatment of PEx to slow lung function decline. This has been credited with improvements in survival for people with CF (PwCF), but there is no consensus on the optimal approach to PEx management. BEAT-CF (Bayesian evidence-adaptive treatment of CF) was established to build an evidence-informed knowledge base for CF management. The BEAT-CF causal model is a directed acyclic graph (DAG) and Bayesian network (BN) for PEx that aims to inform the design and analysis of clinical trials comparing the effectiveness of alternative approaches to PEx management. The causal model describes relationships between background risk factors, treatments, and pathogen colonisation of the airways that affect the outcome of an individual PEx episode. The key factors, outcomes, and causal relationships were elicited from CF clinical experts and together represent current expert understanding of the pathophysiology of a PEx episode, guiding the design of data collection and studies and enabling causal inference. Here, we present the DAG that documents this understanding, along with the processes used in its development, providing transparency around our trial design and study processes, as well as a reusable framework for others.
\end{abstract}

\section{Introduction}

Cystic fibrosis (CF) is an inherited disorder which affects multiple organ systems. Premature death is usually caused by respiratory failure. Loss of lung function occurs progressively, punctuated by acute pulmonary exacerbations (PEx) in which abrupt declines in lung function are not fully recovered. These PEx are inflammatory in nature, and generally considered to be driven by infection of the lower airways.

A cornerstone of CF management in the past half century has been the aggressive treatment of PEx to try to slow the decline of lung function. This has been credited with gains in survival for people with CF (PwCF), but there is no consensus on the optimal approach to PEx management. Therapy is typically multi-modal, comprising the use of broad-spectrum antibiotics, airway clearance therapies, and occasionally the use of immune modulating agents. Surveys of CF clinicians suggest that a wide variety of antibiotic therapies are used in the management of CF exacerbations across Australian CF centres~\citep{Currie2021CFExacerbations}, consistent with the heterogeneity which has been documented in CF cohorts elsewhere~\citep{West+2017Standardized,Cogen+2017Characterization,CogenQuon2024Update}. While numerous trials and observational studies have attempted to define an optimal approach to PEx therapy, such studies have been relatively small and mostly inconclusive~\citep{SANDERS2017592,Cogen+2017Characterization}. Heterogeneity in the interventions studies and the outcomes assessed has complicated attempts to synthesise the evidence. \ywComment{[YW: Would be good land on the need of clinical trials to generate better evidence.]}

BEAT-CF (Bayesian evidence-adaptive treatment of CF) was established in an attempt to build an evidence-informed knowledge base for CF management, addressing the limitations of earlier efforts through scale, patient-centredness, harmonised data capture, and iterative knowledge generation using Bayesian causal inference. In the first instance, BEAT-CF is focussed on the management of PEx, although it intends to apply this approach to other aspects of CF care.

The BEAT-CF causal model for PEx described here was developed to inform the design and analysis of clinical trials to determine the comparative effectiveness of alternative approaches to PEx management. The causal model is a directed acyclic graph (DAG) that represents the causal relationships between background risk factors, treatments and pathogen colonisation of the airways that affect the outcome of an individual PEx episode. The key factors, outcomes and causal relationships were elicited from CF clinical experts and together represent a current expert understanding of the pathophysiology of a PEx episode. The direct effects of alternative PEx management approaches vary, but can all be described in this single framework, and doing so makes their mechanistic effects on downstream outcomes explicit. The DAG therefore enables causal inference~\citep{hernanCausalInferenceWhat2020} --- causal effects of PEx management on PEx outcomes can be estimated using the DAG to identify the likely necessary, sufficient and prohibited adjustments to statistical analyses. In turn, this identifies the observations and measurements that are needed to make these adjustments, directly guiding the design of data collection.

Recently, CF transmembrane conductance regulator (CFTR) modulator therapy has greatly reduced the frequency of PEx. Modulators may have been responsible for most of the 60-70\% decline in PEx in the United States since the turn of 2020, following the approval of Trikafta for patients older than 12 years~\citep{DwightMarshall2021CFTR,CogenQuon2024Update}. However, PEx may still occur, and not all PwCF are eligible for modulator therapy. The consequences of PEx remain severe. A recent study indicates that PEx outcomes may be improved by modulator therapy~\citep{Stone+2025Effect}, however others show similar outcomes with or without its use~\citep{Flume+2018Recovery,McElvaney+2023Impact,Fathima2501349}. The BEAT-CF causal model does not model CFTRs, CFTR modulator therapies or the mechanisms by which they may alter the course of an individual PEx, but as a framework, supports the extension to these and other mechanisms and their associated interventions.

More generally, in the absence of causal DAGs, researchers often leave their causal assumptions implicit or only informally described. The use of causal DAGs in epidemiology and health~\citep{Tennant+2021Use} helps ensure these assumptions are clear and formal. Even still, researchers rarely describe an explicit process that provides confidence that the resulting DAG represents a shared understanding and reasonable assumptions, often without explanation of why the assumptions in the DAG were made. Here, we describe our causal DAG development process for the BEAT-CF causal model (Section~\ref{sec:methods}), and describe the structure (and thereby assumptions) of the model in detail (Section~\ref{sec:results}). We also describe the steps we took to validate the structure of the model, including both qualitative and quantitative validation (Section~\ref{sec:validation}). While uncertainties will always remain, this should provide greater confidence that the BEAT-CF causal model can serve as a useful knowledge base for understanding CF management and the design of observational and trial studies.

\section{Methods} \label{sec:methods}


The purpose of the BEAT-CF causal model is to capture the basic features of PEx onset, pathophysiology and potential longer term consequences, in a way that can be easily extended to support the needs of specific observational and trial studies. Out of scope for the model were detailed interventions and their mechanisms, with extension to these being left for specific studies. Expert information was used to create the model, with elicitation workshops and sessions playing a key role during the development process. The model was originally envisioned as a causal BN, with the DAG (i.e., the BN structure) being developed first and the parameterisation to have followed as an integral part of the causal model. Eventually, the choice was made to shift the focus to the causal DAG, with parameterisation instead serving solely to validate the structure, since the DAG alone would suffice to guide any later database design, statistical analysis and most other research.


Development of the BEAT-CF causal model occurred between 2017 and 2019, with significant inputs into the modelling work occurring via a series of elicitation sessions. Table~\ref{tab:revisions} provides a summary of the changes to the model over time, along with the timing and focus of the main workshops.

The advisory expert developed an initial straw model, primarily as a causal model, in mid 2017. This was used to gain an initial understanding of PEx pathophysiology, its potential complexity and the likely relationships amongst key variables. The straw model was discussed with the modelling team and initial work was done to take structures and patterns from the straw model and incorporate them into a knowledge model (at the time called the conceptual model). To prepare for elicitation, the model variables were divided into 4 main domains common to many clinical models: treatments, background factors (including potential subgroups), endpoints and processes. It was decided that the first 3 domains required input and oversight by a broader set of domain experts, and would therefore be the focus of an expert workshop.

The first workshop was held in Melbourne in August 2017, held as part of the Australian CF Conference. The conference provided a rare opportunity in which many key experts were available in one location, leading to approximately 30 experts participating in the workshop. Participant areas of expertise included the management of children and adults with CF, the science and pathophysiology of CF, and lived experience of CF. The workshop focused on selecting variables for each of the 3 domains, using a method that shares elements with the Delphi and nominal group techniques. Elicitation was divided into 2 rounds, the first focused on generating potential variables and the second focused on discussion of the aggregated results, as well as clustering into groups that would better fit the model resolution.

In the first round, participants worked independently (with no discussion) and were asked to focus on one of the four variable domains (for example, treatments). On the provided handouts, they listed variables they considered important in that domain along with an importance ranking on a scale of 1-5 (1 labelled `not at all important' and 5 `essential') and a rating of their certainty (1 labelled `very unsure' and 5 `very sure'). All responses were anonymous and optional, with the expectation (but not requirement) that participants would opt out of domains outside their expertise. There were 12 responses from experts in the `treatments' domain, 14 responses in the `endpoints' domain and 12 responses in the `subgroups' domain. Responses contained between 3 and 22 suggested items. These were aggregated into lists for each domain in an Excel spreadsheet during the workshop, and assigned a preliminary grouping by a member of the team and the advisory expert. The aggregated lists and groupings were then viewed live, and all participants were invited to discuss and identify items that were considered particularly important, as well as to review the assigned groups. Due to time constraints, importance rankings were only assigned to the variables identified as key and the ranks were agreed to as a group. The resultant aggregate tables are shown in Appendix~\ref{app:agg-responses}.


After the workshop, in September, the modellers grouped and simplified the lists further and decided how the key domains and variables should be represented in the BN. The template was discussed with the broader team in December along with variable selection based on the workshop results. A first concrete version was drafted in late February. This version contained key treatments, subgroups and endpoints as well as some of the most important colonisations (see Appendix~\ref{app:feb2018model}).

Further changes were made to the model structure, including an expansion of the supported endpoints and types of colonisation, in preparation for its qualitative parameterisation towards the end of 2018. Qualitative parameterisation is the process of providing parameters that capture the quality of the local relationships (specifically, the direction of causation, the relative impact of parents and potentially the approximate order of magnitude of effects), without regard to numerical accuracy. The purpose here was twofold: 1) to validate the {\em local} structure of the model by examining whether the conditional probability questions made sense to the experts (i.e., made appropriate direct causal assertions); 2) to validate the {\em global} structure of the model (in preparation for later quantitative statistical analysis) by seeing if the model was capable of exhibiting the global qualitative behaviours that experts expected.

This was followed by a walk-through validation of the model in April 2019. Several scenarios were presented (an example scenario is shown in Appendix~\ref{app:scenario-validation}) and experts were asked to indicate whether the inferred probabilities in that scenario appeared acceptable, or otherwise what probabilities they might expect to see. The walk-through led to minor changes to the structure that the team subsequently addressed. The team did not make adjustments to the parameterisation, as the parameterisation was not further required.

The model was largely finalised in July 2019, with some small adjustments also made in late 2024 and 2025 to align with subsequent trial and data analysis DAGs derived from this model. This consisted of reintroducing `Impaired Mucociliary Clearance' and related treatments, renaming it to `Abnormal Mucus and Clearance' to broaden the scope of the node, adding a direct arrow from `Abnormal Mucus and Clearance' to `Inflammation', and explicitly including `Lung Function'. The final model is described below.

\begin{table}[]
    \centering
    \includegraphics[width=0.9\linewidth]{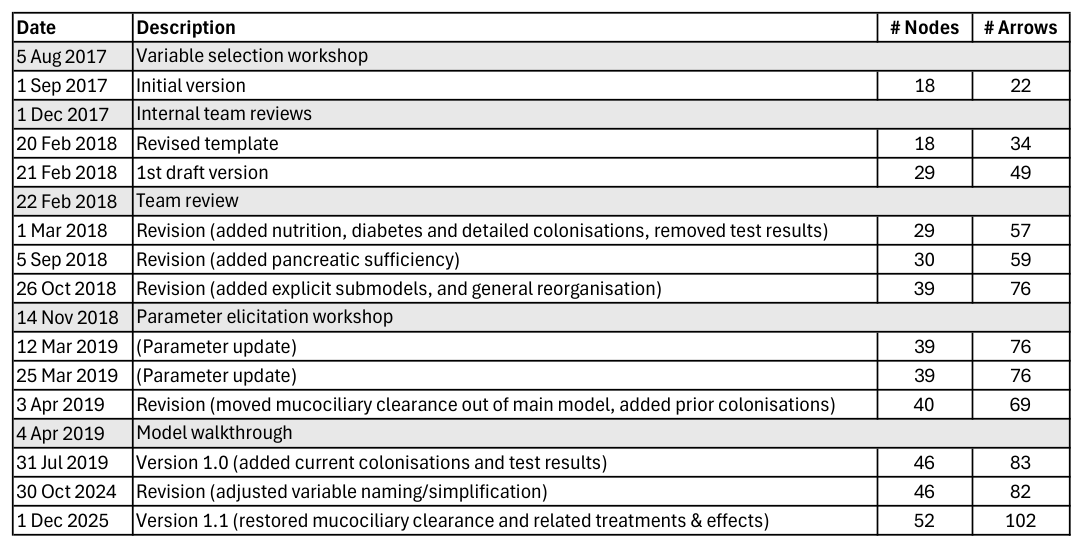}
    \caption{Summary of revisions to the BEAT-CF causal model. Grey rows indicate key workshops and team reviews. The `\# Nodes' and `\# Arrows' columns show the number of nodes and arrows in the model after the given revision.}
    \label{tab:revisions}
\end{table}

\ywComment{[YW: For the proper journal submission, interested in re-organising this via the standard reporting guide. Also, needing to avoid defining experts as research participants - to avoid ethics issue.]}

\section{Results: BEAT-CF Causal Model} \label{sec:results}

\begin{figure}
    \centering
    \includegraphics[width=0.5\linewidth]{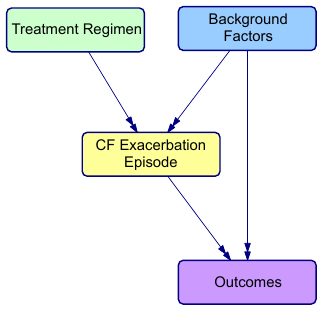}
    \caption{A high-level view of the BEAT-CF causal model, showing the 4 domains and the causal influences between them.}
    \label{fig:high-level}
\end{figure}

There are 4 domains in the BEAT-CF causal model: background factors, treatment regimen, CF exacerbation episode and outcomes. The causal influences between these domains are shown in Figure~\ref{fig:high-level}, and the variables defined within each is provided in Appendix~\ref{app:var-dictionary}. This is similar to a common arrangement for diagnostic medical BNs in which a tier of background variables precedes a tier of disease variables and then a tier of symptoms or test results (for example, see~\citealp{Arora+2019Bayesian}), but also includes treatments. The causal model only considers treatments that affect the exacerbation episode (excluding treatments for symptom management), while background factors may contribute to both the nature of the episode and the outcomes. Domains that were excluded include the epidemiology around infections (including seasonality, transmission and exposure), long-term progression (the model focuses on single exacerbation events) and most tests (such as imaging and blood tests).

\subsection{CF exacerbation}

\begin{figure}[h!]
    \centering
    \includegraphics[width=0.95\textwidth]{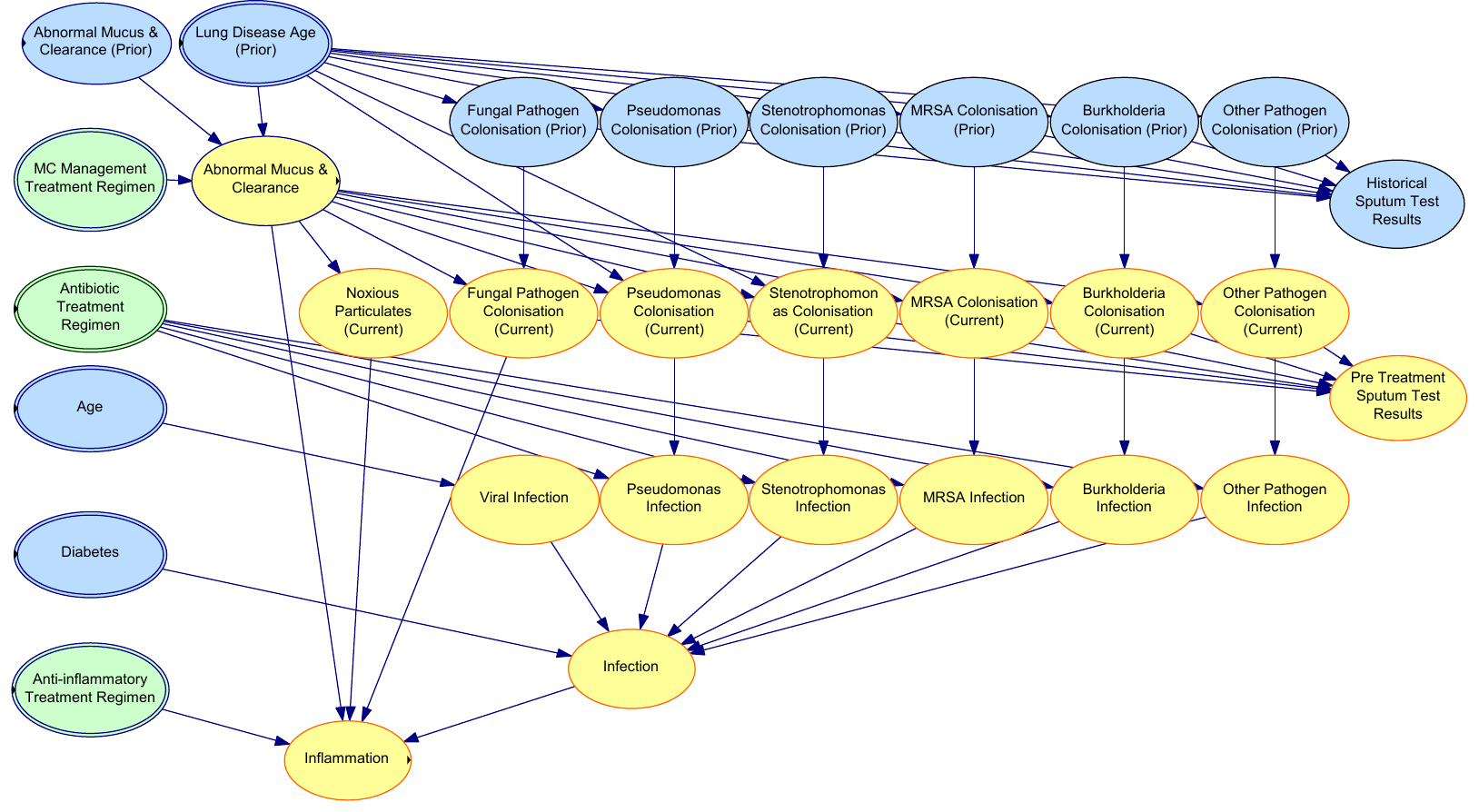}
    \caption{The CF exacerbation submodel showing the core pathophysiological process in an exacerbation event, including abnormal mucus and mucociliary clearance, colonisation, infection and inflammation, as well as key modifying factors.}
    \label{fig:exacerbation}
\end{figure}

The CF exacerbation submodel (Figure~\ref{fig:exacerbation}) depicts the core process involved in the exacerbation event. The submodel is centred around airway colonisation by key groups of respiratory pathogens of concern. The list of pathogens includes key bacterial species, in particular \var{Pseudomonas} aeruginosa, \var{Stenotrophomonas} maltophilia, \var{methicillin-resistant \textit{Staphylococcus aureus}} (MRSA) and \var{Burkholderia} cepacia and \var{fungal pathogens} (such as Aspergillus fumigatus), which are grouped together.\footnote{Phrases in bold refer (either directly or in paraphrase) to specific nodes in the BEAT-CF causal model.} The key bacterial genera and species were kept separate as they were considered to have particular implications for antibiotic treatment during a CF exacerbation and outcomes, however all such colonisations have the same causal structure (with different parameters). Other colonising pathogens not explicitly listed are grouped together into an \var{other pathogen} group. Pathogen colonisations persist over time, and the development and consequences of the exacerbation strongly depend on the state of these prior colonisations. Hence, the submodel includes explicit nodes for these (nodes in blue marked \var{(Prior)}). Prior colonisations can be influenced by an array of individual background factors,
but all of which are assumed to be mediated by \var{lung disease age} (as it was at a prior infection). For any given state of \var{lung disease age}, the model currently assumes a fixed probability that any of the specific pathogen colonisations may be present (which can be updated if more evidence becomes available).

Evidence of \var{prior colonisations}, combined with \var{abnormal mucus and clearance} in the airways, influences the probability of \var{current colonisations}, that is, colonisations at the time of the exacerbation. These too can be tested for, and the results entered as evidence into the model via \var{pre-treatment sputum test results}. In the absence of more timely evidence, \var{historical sputum test results} provide strong differential evidence about the current colonisation status, because prior colonisation nodes represent the presence of colonisation {\em at the time} of the \var{historical sputum test results}.

Any of the colonisations may lead to an \var{infection} wherein the microbes cause tissue damage and provoke an inflammatory immune response. \var{Abnormal mucus and clearance} may trigger \var{inflammation} directly, for example, due to airway obstruction leading to inflammatory signalling, or due to trapped \var{noxious particulates}, or otherwise by making pathogen colonisations more likely. While it would be possible to collapse \var{abnormal mucus}, \var{colonisation}, \var{infection} and \var{inflammation} together as a single mediator, this would not serve the aim of the model, which is to represent the effectiveness of treatments for exacerbations. Colonisation without infection does not need treatment, but sputum test results do not distinguish colonisation from infection. Colonisation may lead to \var{infection} treatable with \var{antibiotics} and \var{infection} may lead to \var{inflammation} treatable with an \var{anti-inflammatory} agent. Issues with mucus may be managed with \var{mucociliary treatments} but not \var{antibiotics}. Hence, all 4 features ---  abnormal mucus and clearance, colonisation, infection and inflammation --- play a different role in mediating the effect of an exacerbation treatment on the outcome.

A virus may also cause an infection, with similar consequences to other infections. However, \var{viral infections} differ in that they are not treatable with \var{antibiotics} or \var{mucociliary management}, do not generally result in persistent asymptomatic colonisation, and are more likely to arise independently of the state of \var{lung disease age}. They are, however, still dependent on the individual's \var{age}. While \var{age} may also increase the chance of other infections independent of \var{lung disease age}, existing colonisations dominate for an individual with CF. 
Any \var{infection} is made more likely by the presence of \var{diabetes} via several mediators not included in the causal DAG (including impaired immune function, high blood sugar, and others).

Alongside current \var{bacterial colonisation} and \var{noxious particulates}, \var{fungal colonisation} may also be present and provoke or aggravate \var{inflammation}. While \var{fungal colonisation} can rarely result in \var{infection} in those with compromised immunity, more typically colonisation results in inflammation without infection (i.e., allergic bronchopulmonary aspergillosis). Hence, we only included the direct arrow from \var{fungal colonisation} to \var{inflammation}, which may or may not have infection as an implicit mediator. While the effects of mucus, particulates, fungi, viruses and diabetes are not affected by treatment with antibiotics, all can lead to \var{inflammation}, which can be affected by treatment with \var{anti-inflammatories}.

\subsection{Background factors}

\begin{figure}[h!]
    \centering
    \includegraphics[width=0.8\textwidth]{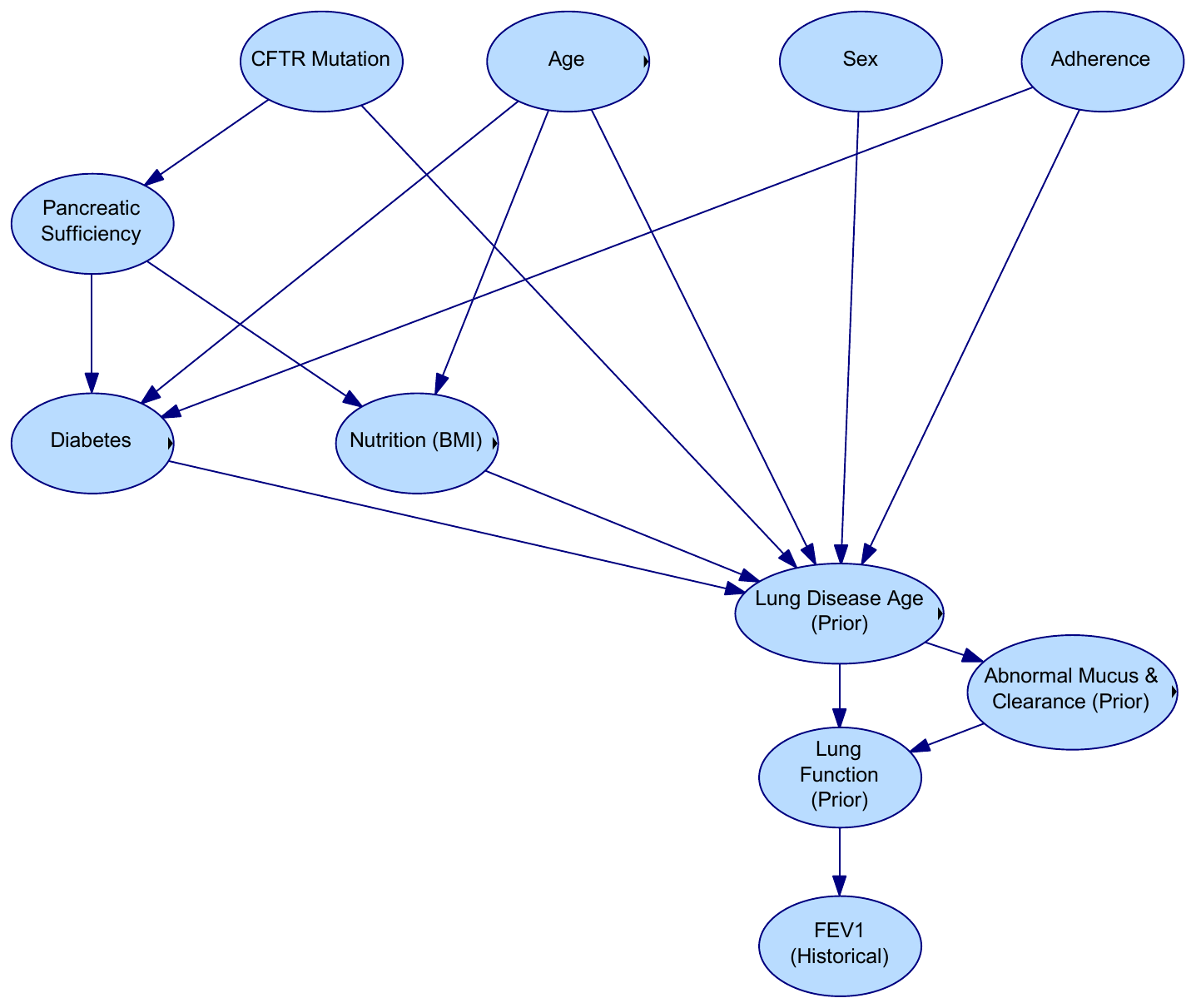}
    \caption{Background factors that affect the incidence and nature of an exacerbation}
    \label{fig:background}
\end{figure}

The progression of the exacerbation will be affected by characteristics of the individual at the time of exacerbation (see Figure~\ref{fig:background}). All background variables ultimately influence the exacerbation via their effect on \var{lung disease age} prior to infection, which is a surrogate for the progression of disease in the individual's lungs. Together with \var{abnormal mucus and clearance}, this largely determines the functional capacity of the individual's lungs (\var{lung function}) and can be measured by the forced expiratory volume in 1 second \var{FEV1}. There are several relationships of note between the background variables. The \var{CFTR mutation} class can affect \var{pancreatic sufficiency}, which in turn affects the \var{nutrition} status of the individual along with the chance of having \var{diabetes}. \var{Diabetes} indicates both whether the individual has diabetes, and whether or not it is well managed. 
An individual's \var{adherence} to treatments strongly influences how well their disease condition is (historically and currently) managed, which can include both \var{diabetes} and the condition of their lungs as indicated by \var{lung disease age}.

Returning to Figure~\ref{fig:exacerbation}, \var{lung disease age} primarily affects the chance of having each type of colonisation at the time of earlier test results, or at the beginning of the current exacerbation. \var{Age} and \var{diabetes} also influence the progression of an exacerbation independent of \var{lung disease age}, with \var{age} affecting the chance of having a \var{viral infection}, and \var{diabetes} affecting the probability of a colonisation becoming \var{infection}. 

\subsection{Treatments}

\begin{figure}[h!]
    \centering
    \includegraphics[width=0.9\textwidth]{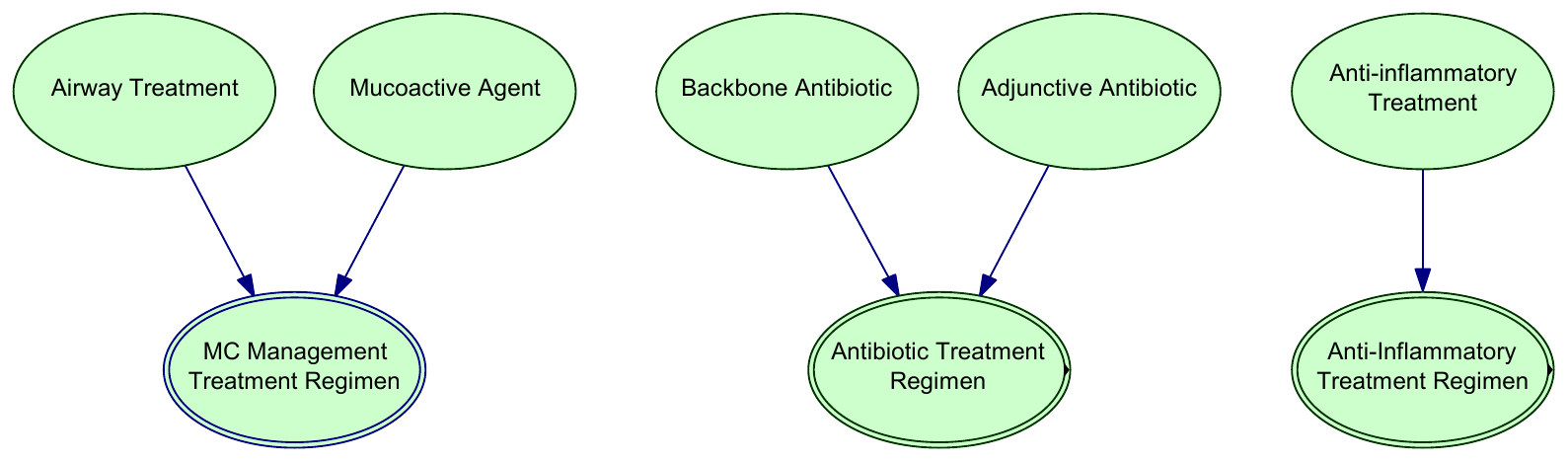}
    \caption{Treatments for a CF exacerbation included in the BEAT-CF causal model}
    \label{fig:treatments}
\end{figure}

The severity of an exacerbation can be moderated by treatment. Figure~\ref{fig:treatments} shows the three major groups of treatments available in the model, \var{mucociliary management}, \var{antibiotics} and \var{anti-inflammatories}. Mucociliary management is divided into the \var{mucoactive agent} employed (such as dornase alfa or hypertonic saline) and the frequency of \var{airway treatments} (such as chest physiotherapy). Antibiotics are divided into \var{backbone} and \var{adjunctive}, which, together with frequency and dosage, determines the \var{antibiotic treatment regimen}. Similarly, an anti-inflammatory at a particular frequency and dosage determines the \var{anti-inflammatory treatment regimen}. As described earlier, these three treatment regimens reduce the chance of abnormal mucus and clearance, infection and inflammation in the exacerbation model.

\subsection{Outcomes} \label{sec:outcomes}

\begin{figure}[h!]
    \centering
    \includegraphics[width=0.9\textwidth]{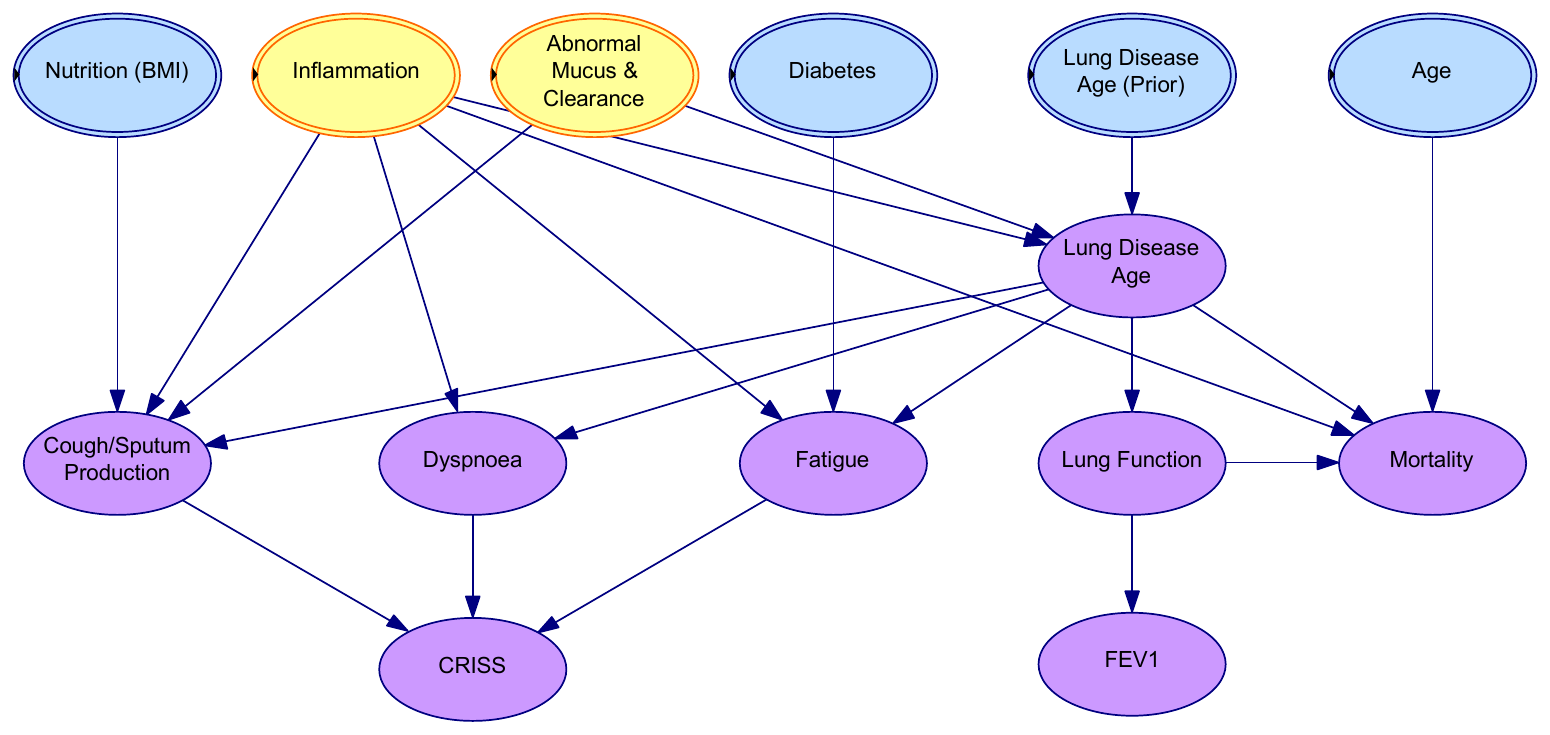}
    \caption{Transient and persistent consequences of a CF exacerbation. Longer term outcomes include changes to lung function and mortality}
    \label{fig:outcomes}
\end{figure}

Taking into account the effect of treatments and background factors, the exacerbation can lead to outcomes (see Figure~\ref{fig:outcomes}) that may persist or worsen over time. This may include a persistent change to \var{lung disease age} and \var{lung function}, which, along with \var{inflammation} and \var{abnormal mucus and clearance}, may have consequences for \var{cough and sputum production}, \var{dyspnoea} and \var{fatigue}. Again, \var{lung function} can be measured directly with \var{FEV1}, while the other 3 outcomes can be measured by the Chronic Respiratory Infection Symptom Score \var{CRISS} assessment tool. Changes to \var{lung disease age}, as well as declining \var{lung function} or severe \var{inflammation}, all increase the chance of \var{mortality}. Various background factors can directly modify outcomes. In particular, \var{age} affects \var{mortality}, \var{nutrition} affects \var{cough and sputum production} and \var{diabetes} affects \var{fatigue}. All other background, treatment and exacerbation factors are otherwise mediated by \var{lung disease age} and \var{inflammation}.


\subsection{Validation} \label{sec:validation}

The purpose of the model was to inform later statistical models for trial analyses, hence there was no need to parameterise the BN structure or perform a quantitative validation. However, there were still benefits to validating the structure itself. The structure was validated primarily via three means: 1) model structure walkthroughs at two team review sessions, 2) a parameterisation workshop that tested local structure and 3) scenario validations against a qualitatively parameterised version of the BN.

The first team review in mid September 2017 (see Table~\ref{tab:revisions}) concluded that the high-level structure of the model was appropriate, and led primarily to refinements in which arrows needed to be added or removed, particularly those involving background factor impacts on exacerbations. The second team review involved a walkthrough of the structure and discussed specifically how variables were related to their parent causes. This resulted in the addition of several arrows, as well as several new nodes including nutrition, fungal pathogen colonisations and noxious particulates.

A parameterisation workshop was conducted in November 2018. This tested local causal structures (both arrows and relationship types) by asking the two clinicians within the team to estimate best case and worst case distributions for each node, along with the causal strength of each parent relative to the node's other parents. No structural issues were raised by the experts when providing answers to the questions on causal strength, and no requests were made to include other possible causes. This provided evidence that the arrows were directed correctly, and that the combined set of parents for each node (i.e., the direct causes) were suitable and likely sufficient to represent the key causes of the node.

Finally, a model walk-through of the parameterised BN was conducted again in April 2019 with both clinicians. The validation suggested that local influences were in the right direction and had the right relative impact in the case of direct causes. There was also no indication that there were erroneous influences between variables, and only two additional causes that were added: \var{adherence} and \var{sex}. These results further suggested the structure was robust. However, the absolute impacts of parent nodes were sometimes weaker than expected, and the sizes of the probability changes over longer paths were often much weaker than expected, even if in the right direction. These issues were not pursued, as they pertained only to the parameterisation. By contrast, no further issues were raised about the structure, suggesting it was causally sound and sufficient for its future purpose.

A final review in 2025 identified the need to reintroduce and expand the mucociliary clearance mechanism. This had been removed from an earlier version of the model, due to the expectation later uses of the model would not cover treatments such as mucociliary management. It was decided during review that the more complete picture was appropriate for the reference BEAT-CF causal model.

\section{Discussion}

Causal DAGs are increasingly used to document the causal assumptions made in observational or interventional studies that attempt causal inference~\citep{Tennant+2021Use}. The DAG serves as a knowledge model that will support our future studies. In particular, by publishing this model prior to using it in our own trials, we improve the transparency of our modelling assumptions, providing a record of our modelling decisions and any changes. We also offer it as a reusable framework for others who may perform similar studies. The causal DAG will be of particular benefit for those who use data recorded in the BEAT-CF registry, since the design of the registry's data structures was informed by the same causal model. \ywComment{[YW: Highlight the intention to inform trial design here.]}

We believe the BEAT-CF causal model captures the key relationships that are relevant for causal effect estimation, but do not suggest that the model is exhaustive or mechanistically precise. We expect that the causal arrows that have been omitted are likely to be weak enough to make their omission acceptable --- that is, to cause only minor bias to the main causal estimations --- and that the depth of mechanistic detail is light but sufficient to capture the kinds of interventions we might want to test. Specific studies will almost certainly need to add further nodes and arrows, but we hope this framework greatly eases the work involved and provides greater confidence in the final causal model.

In applications of the BEAT-CF causal model, we expect the structure itself to be adapted to suit the purpose of the application. We expect this will typically involve simplifying the structure --- focusing just on what is required for the application. This might involve collapsing multiple nodes into one. For example, if the type of colonisation were not relevant to estimating the impact of therapies, the colonisation nodes could be collapsed. Or if new anti-inflammatories were being trialled, colonisation and infection could be collapsed. However, in other cases, such as applications involving mediation analysis or specific subgroups or background factors, we expect that extra detail will need to be added, since the model is not detailed enough to provide good insight into many important mediating mechanisms (such as inflammatory processes) or background factors (such as vaccination status and seasonality). Also, because our primary use case was randomised trials, we have not considered factors that influence treatment decisions. Those planning observational studies will also need to consider these as potential confounders. It nonetheless would serve as a useful starting point.

An application of the BEAT-CF causal model already exists. The model developed here supported work on an evaluation measure for trials based on outcomes for people with CF~\citep{McLeod+2021Novel}. That work adopted a similar expert-driven DAG development approach to expand notably on the outcomes submodel described in Section~\ref{sec:outcomes}. By developing this more detailed model of the outcomes in close partnership with people with lived experience, we could identify causally relevant endpoints, reduce redundant endpoint measurements and improve the chances of orthogonality (independence) amongst the endpoints chosen, creating a more robust and comprehensive measure of outcomes that matter to people with CF. The detailed outcome models that came from this work can also be used as a replacement for the simpler outcome submodel presented here where the finer details are required.

We made use of qualitative parameterisation solely to validate the model structure. However a more strongly validated parameterisation could be of notable value in designing trial and observational studies, and ultimately serve as a basis for a future clinical decision-support tool based on quantitative prognostication for individuals. In terms of study design, a representative parameterisation can support sensitivity analyses and useful estimates of the strength of individual arrows. This, in turn, provides a much better understanding of {\em which} confounders or backpaths are likely to be ignorable, and which need to be accounted for in the study design in order to avoid significant estimation bias. Similarly, it would allow the impact of potential selection biases to be estimated and explored, providing a better understanding of how to efficiently mitigate these impacts. It could help with estimating effect sizes prior to conducting a trial, and, in turn, the sample sizes that may be required to obtain useful results. In Bayesian analyses, the parameterisation can also be used as a prior to regularise the effect estimates (by placing low probability on unlikely effect sizes), to better handle missing data, and as an informative prior upon which new data can accrue to reflect a more up-to-date understanding of the pathophysiology of interest. While potentially very useful, the benefits of parameterisation must be weighed against the extra effort required to produce usefully representative parameters.

\section{Conclusion}

As \citet{Tennant+2021Use} notes, causal DAGs are often developed without a structured process. We believe we have presented and used here a relatively simple structured process to develop a causal model of CF exacerbations. The process is expert-driven and provides multiple opportunities for structure validation, both via direct feedback from experts and the more stringent tests that qualitative parameterisation provides. While there is room to improve both the quality of the validation and the efficiency of the process, improvements which we are actively exploring through our other work, we believe that this process can be adopted at low cost for great benefit now.

We plan to extend the causal model here to support specific observational studies, both emulated and real trial questions, and serve as a basis for the development of causal model-driven clinical decision-support tools. A key test of the value of the causal model will be how well it extends to these applications. We also plan to extend the model beyond individual exacerbation events, providing a broader picture of the pathophysiology and consequences of exacerbations for people with CF. 

\section{Acknowledgement}

Funding for this research has been provided from the Australian Government's Medical Research Future Fund (MRFF), GNT1152376. The authors are grateful for the input provided
by domain experts and attendees at Australian CF Conference 2017, as well as the input and support from the community for this approach via our Consumer Reference Group (BEAT CF CRG). The models described in this paper were created using the GeNIe Modeler, available free of charge for academic research and teaching use from BayesFusion, LLC, https://www.bayesfusion.com/.

\bibliography{citations}
\bibliographystyle{apalike}






\newpage
\appendix
\section{Aggregated responses from August 2017 workshop: treatment domains, subgroups and endpoints} \label{app:agg-responses}

\subsection{Treatment}

\begin{table}[h!]
\begin{tabular}{ll}
\textbf{Treatment   domains} & \textbf{Importance} \\
\hline
Backbone antibiotics & 5 \\
Airway clearance & 5 \\
Psychology & 5 \\
Adjunctive Antibiotics & 5 \\
Nutrition & 3 \\
Mucoactive agents & 2 \\
Exercise & 2 \\
Anti-inflammatory & 1
\end{tabular}
\end{table}

\subsection{Background factors/subgroups}

\begin{table}[h!]
\begin{tabular}{llll}
\textbf{Name} & \textbf{Type} & \textbf{Influences treatment?} & \textbf{Importance} \\
\hline
renal impairment & comorbidity & y & 5 \\
liver impairment & comorbidity & y & 5 \\
drug allergies/ reactions & comorbidity & y & 5 \\
pregnancy & comorbidity & y & 5 \\
pseudomonas/cepacia & micro & y & 5 \\
other resistant organisms & micro & y & 5 \\
hearing impairment & comorbidity & y & 5 \\
patient/physician reported previous non-response & other & y & 4 \\
disease severity of FEV1 & severity & y & 4 \\
stenotrophomonas & micro & y & 3 \\
age & demographic & y & 2 \\
staph aureus & micro & y &  \\
diabetes & comorbidity &  &  \\
ABPA & comorbidity &  &  \\
pancreatic sufficiency & comorbidity &  &  \\
location & demographic &  &  \\
Ethnicity & demographic &  &  \\
first exacerbation & disease stage &  &  \\
recent exacerbation & disease stage &  &  \\
frequency of previous exacerbation & disease stage &  &  \\
historical microbiology & micro &  &  \\
probability of adherance & other &  &  \\
IV access & other &  &  \\
exacerbation severity & severity &  &  \\
viral features & symptoms &  &  \\
haemoptysis & symptoms &  & 
\end{tabular}
\end{table}

\newpage
\subsection{Outcomes/endpoints}
\begin{table}[h!]
\begin{tabular}{lll}
\textbf{Name} & \textbf{Type} & \textbf{Importance} \\
\hline
Exercise tolerance & function & 5 \\
Mortality & function & 5 \\
FEV1 & objective sign & 5 \\
General adverse effects, serious & safty & 5 \\
Fatigue & symptom & 5 \\
Appetite & symptom & 5 \\
CRISS & validated score & 5 \\
Anxiety/ depression scales & validated score & 5 \\
Oxygen saturation & objective sign & 4 \\
Weight & objective sign & 4 \\
General adverse effects, others & safty & 4 \\
Quality of life/well-being score & validated score & 4 \\
CF RSD & validated score & 4 \\
CAT score/ questionnaire & validated score & 4 \\
CRP & biomarker &  \\
Ability to return to school/work & function &  \\
Eradication of infection & micro &  \\
Microbiological response & micro &  \\
Xray & radiology &  \\
CT scan & radiology &  \\
resolution of auscultatory findings & sign &  \\
Cough no longer productive & symptom &  \\
Severity of exacerbation & symptom &  \\
Wonders of sputum & symptom &  \\
Anxiety/ depression & symptom &  \\
Resolution of haemoptysis & symptom &  \\
Sleep quality & symptom &  \\
Number of community exacerbation & treatment &  \\
Time to next exacerbation & treatment &  \\
Time to home/ days in hospital & treatment &  \\
Need/ number of adjunctive therapies & treatment & 
\end{tabular}
\end{table}

\clearpage
\newpage
\section{BEAT-CF Causal Model - February 2018 version} \label{app:feb2018model}

\begin{figure}[h!]
    \centering
    \includegraphics[width=1\linewidth]{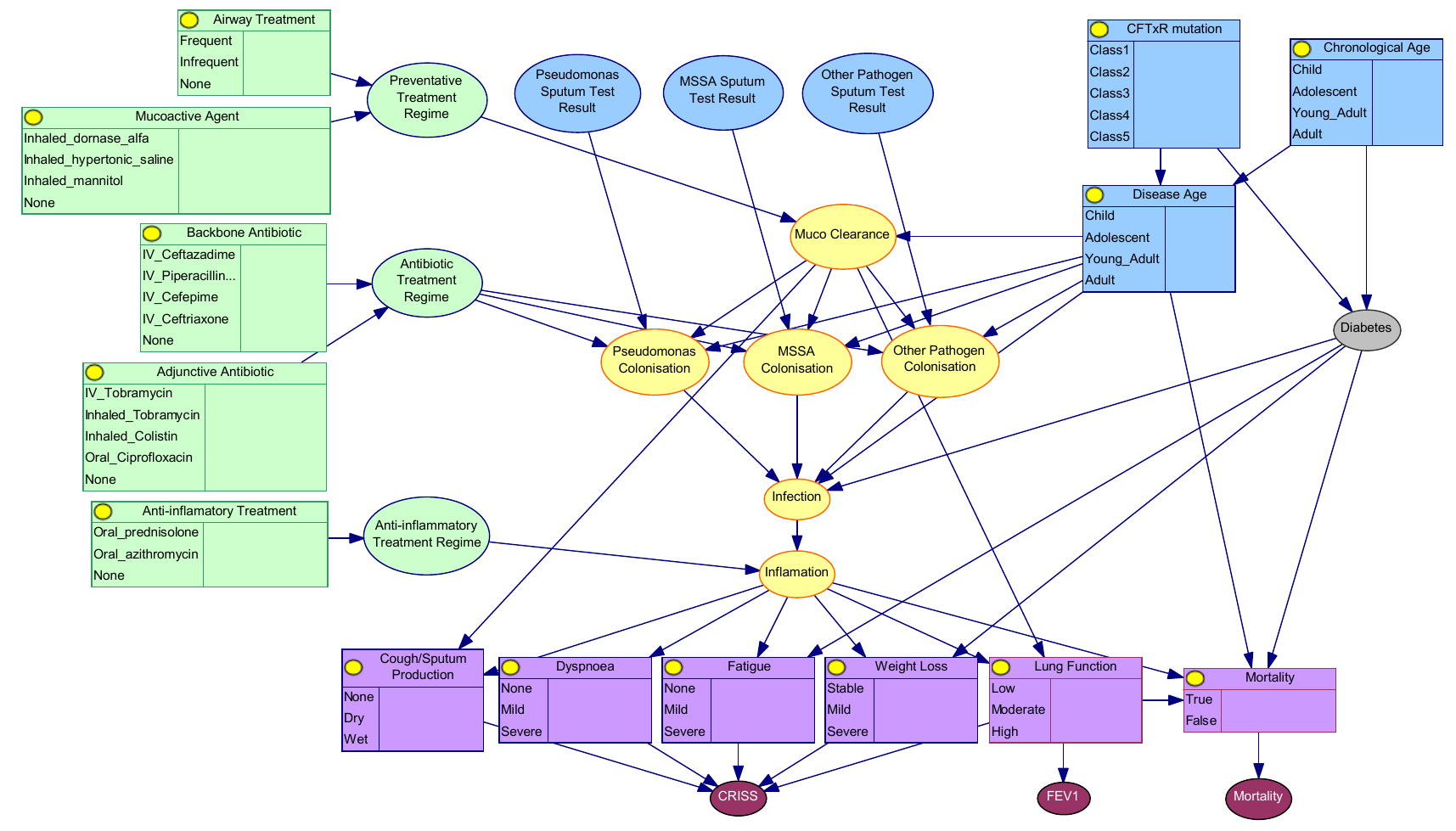}
    \label{fig:placeholder}
\end{figure}

\clearpage
\newpage
\section{Qualitative parameterisation validation example} \label{app:scenario-validation}

\begin{figure}[h!]
    \centering
    \includegraphics[width=0.9\textwidth]{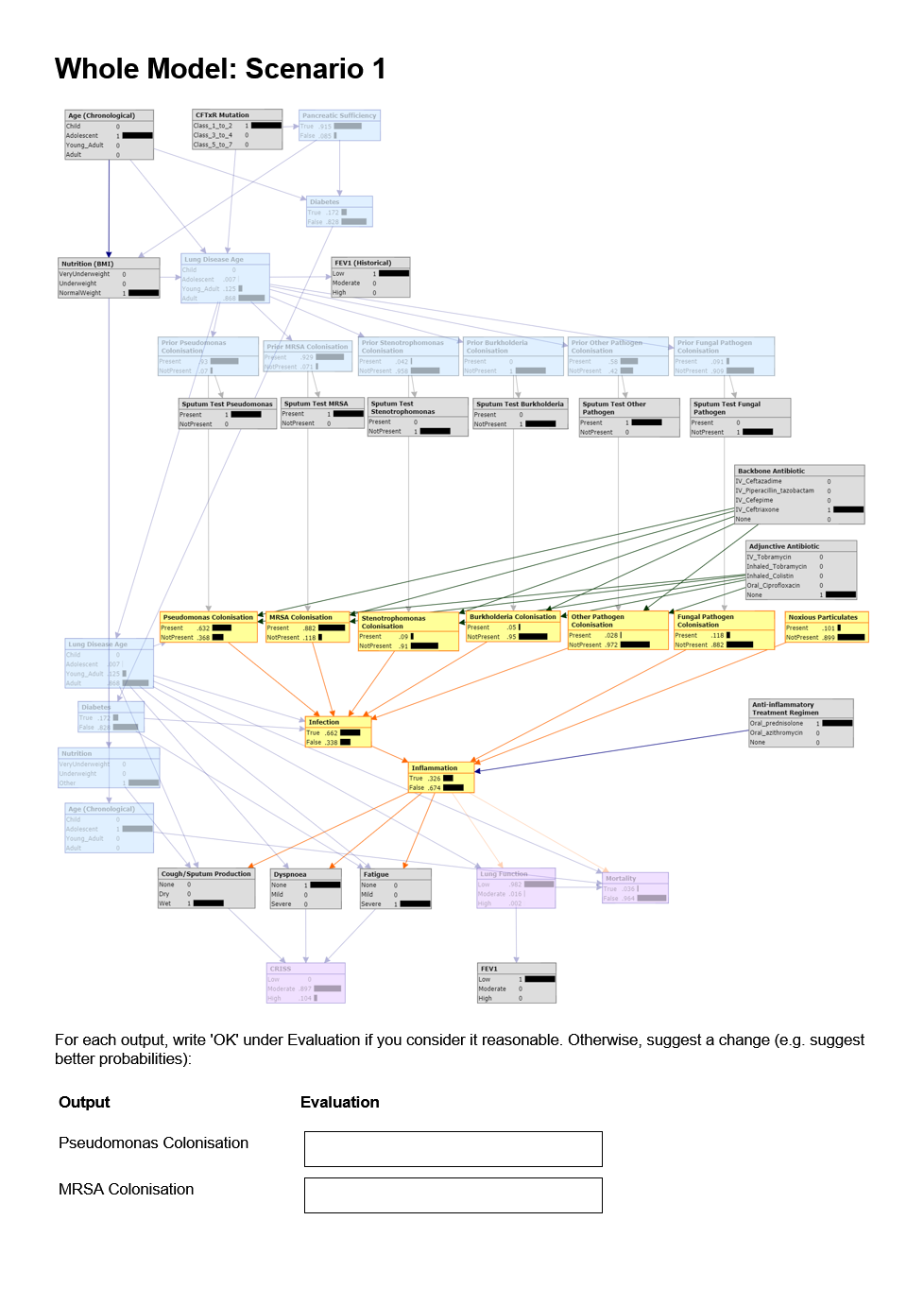}
    \label{fig:scenario-validation}
\end{figure}

\begin{figure}[h!]
    \centering
    \includegraphics[width=\textwidth]{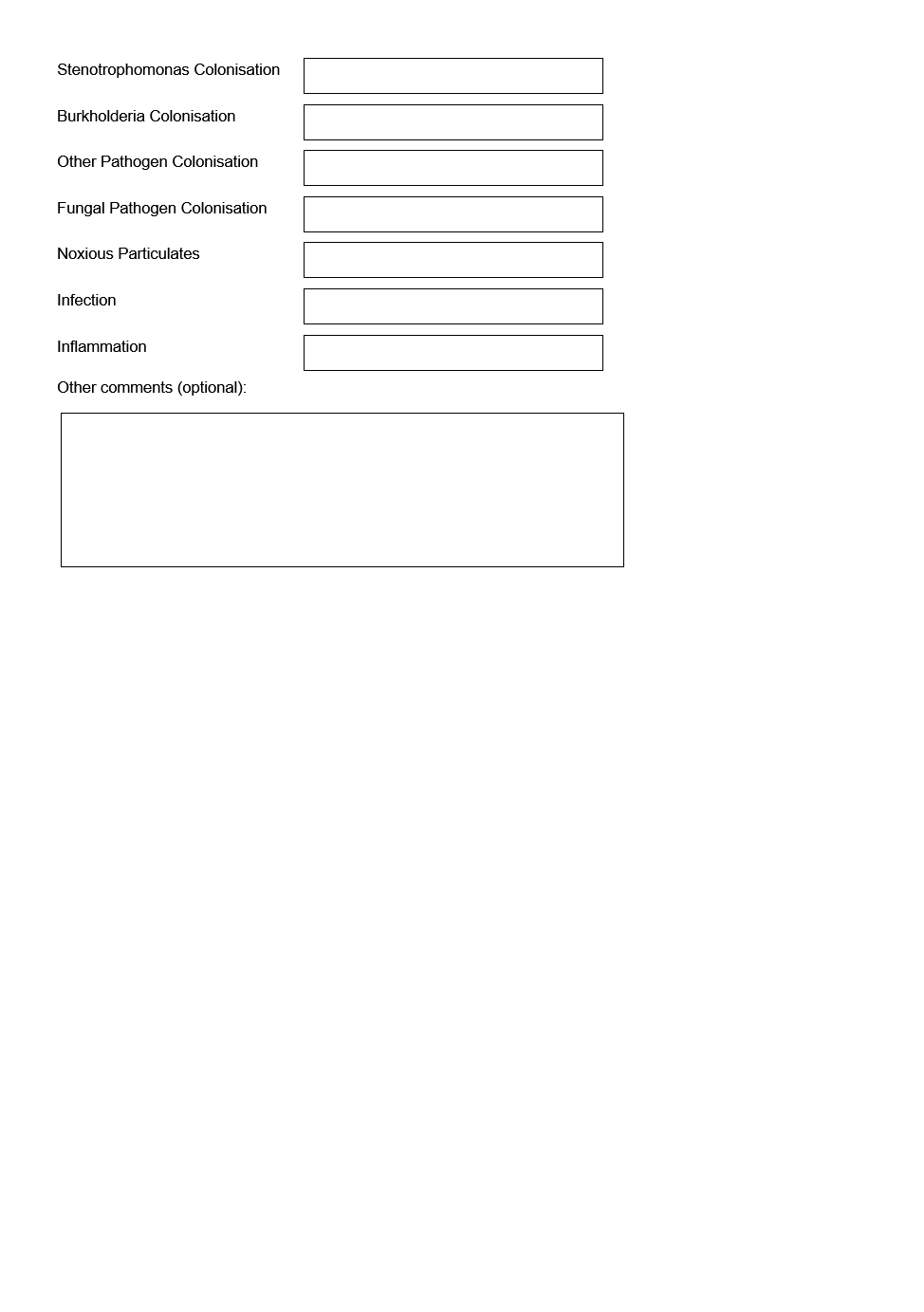}
    \label{fig:scenario-validation2}
\end{figure}

\clearpage
\newpage


\section*{Supplementary material}

# D Variable dictionary

| Treatment Regimen | | | | |
|---|---|---|---|---|
| Variable | Type (Indicative) | Description | Parents | Relationship to Parents |
| MC Management Treatment Regimen | Categorical | The combination of mucoactive agent and airway treatment frequency, targeting issues with | Airway Treatment, Mucoactive Agent | Uniform combination of the two parents (but only one 'none' state) |
| Airway treatment | Ordered: Frequent, Infrequent, None | Physical therapies aimed at improving airway clearance, specified as the frequency with which airway treatments are provided at the time of the | - | N/A |
| Mucoactive Agent | Categorical: Inhaled_dornase_alfa, Inhaled_hypertonic_saline, Inhaled_mannitol, None | Medications used to improve mucus quality and/or mucociliary clearance | - | N/A |
| Antibiotic Treatment Regimen | Categorical | Antibiotic treatments for an exacerbation (types and dosage combination) targeting the propensity to have an infection or eliminating a | Backbone Antibiotics, Adjunct Antibiotics | Uniform combination of the two parents |
| Backbone Antibiotics | Categorical: IV_Ceftazadime, IV_Piperacillin_tazobactam, IV_Cefepime, IV_Ceftriaxone | The main antibiotic class used for treatment of the exacerbation | - | N/A |
| Adjunct Antibiotics | Categorical: IV_Tobramycin, Inhaled_Tobramycin, Inhaled_Colistin, Oral_Ciprofloxacin | An antibiotic that supplements the main antibiotic to improve coverage or treat co-infections for the exacerbation | - | N/A |
| Anti-inflammatory Treatment Regimen | Oral_prednisolone, Oral_azithromycin | Treatments that aim to prevent or reduce inflammation. For consistency with other treatments, the causal model contains two instances of this node (one for class and one for | - | N/A |

| Background Factors | | | | |
|---|---|---|---|---|
| Variable | Type (Indicative) | Description | Parents | Relationship to Parents |
| CFTR Mutation | Categorical: Classes I - VI | Class of CFTR gene mutation in individual with CF, influence CF severity and organ involvement. Since the causal model assumes an individual with CF, there is no 'Normal' class. Class | - | N/A |
| Age (Chronological) | Ordered: Child, Adolescent, Young Adult, Adult | Chronological age category of the individual at the time of the exacerbation | - | N/A |
| Sex | Categorical: Male, Female | The biological sex of the individual having the | - | N/A |
| Adherence | Ordered: Good, Poor | How well the individual follows any prescribed treatments and management, including for CF | - | N/A |
| Pancreatic Sufficiency | Boolean | Whether the pancreas can produce sufficient digestive enzymes. Insufficiency increases | CFTR Mutation | Higher classes of CFTR mutation are more likely to lead to pancreatic |
| Nutrition (BMI) | Ordered: Underweight, Normal, Overweight, Obese | BMI of individual, calculated from weight and height, and adjusted for sex | Pancreatic Sufficiency, Age | Problems with pancreatic sufficiency can lead to malnourishment. BMI |
| Diabetes | Ordered: None, Well Managed, Present | Presence of diabetes (CF-related or otherwise), which affects infection risk, nutrition and lung disease progression | Pancreatic Sufficiency, Age, Adherence | Older age and pancreatic insufficiency increase the chance of diabetes, while adherence to treatments improves the chance of |
| Lung Disease Age (Prior) | Ordered: Child, Adolescent, Young Adult, Adult | The state of disease progression of the lungs at a prior infection | CFTR Mutation, Age, Nutrition (BMI), Diabetes, Sex, Adherence | Older age, higher classes of CFTR mutation, as well as problems with nutrition, diabetes or adherence all increase the chance of higher lung disease ages. Females generally have |
| Abnormal Mucus & Clearance (Prior) | Ordered: None, Moderate, Severe | Degree to which mucus viscosity, hydration, and ciliary function are impaired, reducing airway | Lung Disease Age (Prior) | Higher lung disease ages increase chance of abnormal mucus and |
| Lung Function (Prior) | Ordered: Low, Moderate, High | Lung function and capacity at a prior infection, acting as a baseline | Lung Disease Age (Prior), Abnormal Mucus & | Increases in both influences impact on lung function |
| FEV1 (Historical) | Ordered: Low, Moderate, High | Historical forced expiratory volume test result, acting as a measurement of the baseline | Lung Function (Prior) | FEV1 is a direct but noisy indicator of lung function |
| Pseudomonas Colonisation (Prior) | Boolean | Presence of Pseudomonas aeruginosa colonisation at a prior infection | Lung Disease Age (Prior) | Higher lung disease age increases chance of colonisation |
| MRSA Colonisation (Prior) | Boolean | Presence of methicillin-resistant Staphylococcus aureus colonisation at a prior infection | Lung Disease Age (Prior) | Higher lung disease age increases chance of colonisation |
| Stenotrophomonas Colonisation (Prior) | Boolean | Presence of Stenotrophomonas maltophilia colonisation at a prior infection | Lung Disease Age (Prior) | Higher lung disease age increases chance of colonisation |
| Burkholderia Colonisation (Prior) | Boolean | Presence of Burkholderia cepacia colonisation at a prior infection | Lung Disease Age (Prior) | Higher lung disease age increases chance of colonisation |

| Variable | Type (Indicative) | Description | Parents | Relationship to Parents |
|---|---|---|---|---|
| Fungal Pathogen Colonisation (Prior) | Boolean | Presence of fungal colonisation (e.g., Aspergillus fumigatus) at a prior infection | Lung Disease Age (Prior) | Higher lung disease age increases chance of colonisation |
| Other Pathogen Colonisation (Prior) | Boolean | Presence of any other pathogen colonisation at a prior infection | Lung Disease Age (Prior) | Higher lung disease age increases chance of colonisation |
| Historical Sputum Test Results | Boolean | Pathogens detected in sputum test results from a prior infection. In practice, there would be one node per colonisation, representing detection of each colonisation | Pseudomonas Colonisation (Prior), Stenotrophomonas Colonisation (Prior), MRSA Colonisation (Prior), Burkholderia Colonisation (Prior), Fungal Pathogen | Sputum test results are a direct but noisy indicator of the presence of each colonisation |

| CF Exacerbation Episode ||||| 
|---|---|---|---|---|
| Variable | Type (Indicative) | Description | Parents | Relationship to Parents |
| Abnormal Mucus & Clearance | Ordered: None, Moderate, Severe | Degree to which mucus viscosity, hydration, and ciliary function are impaired, reducing airway clearance during the current episode | Lung Disease Age (Prior), MC Management Treatment Regimen, Abnormal Mucus & | Inertia (from prior abnormal mucus and clearance). Higher prior lung disease ages increase chance of abnormal mucus and clearance now. |
| Noxious Particulates (Current) | Boolean | Any harmful airborne particles (such as smoke, pollution, dust, etc.) that may irritate the lungs and trigger and inflammatory response | Abnormal Mucus & Clearance | Abnormal mucus and clearance can make it more likely that noxious particles are trapped in the airway |
| Pseudomonas Colonisation (Current) | Boolean | Presence of Pseudomonas aeruginosa colonisation during the current episode | Pseudomonas Colonisation (Prior), Lung Disease Age (Prior), Abnormal Mucus & | Inertia (from prior colonisation). Higher prior lung disease age also increases chance of current colonisation, and abnormal mucus |
| MRSA Colonisation (Current) | Boolean | Presence of methicillin-resistant Staphylococcus aureus colonisation during the current episode | MRSA Colonisation (Prior), Abnormal Mucus & Clearance | Inertia (from prior colonisation). Higher prior lung disease age also increases chance of current colonisation, and abnormal mucus |
| Stenotrophomonas Colonisation (Current) | Boolean | Presence of Stenotrophomonas maltophilia colonisation during the current episode | Stenotrophomonas Colonisation (Prior), Lung Disease Age (Prior), Abnormal Mucus & | Inertia (from prior colonisation). Higher prior lung disease age also increases chance of current colonisation, and abnormal mucus |
| Burkholderia Colonisation (Current) | Boolean | Presence of Burkholderia cepacia colonisation during the current episode | Burkholderia Colonisation (Prior), Abnormal Mucus & Clearance | Inertia (from prior colonisation). Higher prior lung disease age also increases chance of current colonisation, and abnormal mucus |
| Fungal Pathogen Colonisation (Current) | Boolean | Presence of fungal colonisation (e.g., Aspergillus fumigatus) during the current episode | Fungal Pathogen Colonisation (Prior), Abnormal Mucus & Clearance | Inertia (from prior colonisation). Higher prior lung disease age also increases chance of current colonisation, and abnormal mucus |
| Other Pathogen Colonisation (Current) | Boolean | Presence of any other pathogen colonisation during the current episode | Other Pathogen Colonisation (Prior), Abnormal Mucus & Clearance | Inertia (from prior colonisation). Higher prior lung disease age also increases chance of current colonisation, and abnormal mucus |
| Pre Treatment Sputum Test Results | Boolean | Pathogens detected in sputum test results during the current episode. In practice, there would be one node per colonisation, representing detection of each colonisation | Pseudomonas Colonisation (Current), Stenotrophomonas Colonisation (Current), MRSA Colonisation (Current), Burkholderia Colonisation (Current), Other Pathogen Colonisation (Current), Fungal Pathogen Colonisation (Current) | Sputum test results are a direct but noisy indicator of the presence of each colonisation |
| Viral Infection | Boolean | Presence of viral respiratory pathogens, such as RSV, influenza, rhinovirus, etc. | Age | Both younger and older ages are more vulnerable to infection, mostly mediated by a weaker immune |
| Pseudomonas Infection | Boolean | Refers to active tissue damage likely invoking some form of inflammatory response caused by Pseudomonas | Pseudomonas Colonisation (Current), Antibiotic Treatment | Colonisation can lead to infection. Antibiotics can treat and reduce chance of infections |
| MRSA Infection | Boolean | Refers to active tissue damage likely invoking some form of inflammatory response caused by | MRSA Colonisation (Current), Antibiotic | Colonisation can lead to infection. Antibiotics can treat and reduce |

19

| Variable | Type (Indicative) | Description | Parents | Relationship to Parents |
|---|---|---|---|---|
| Stenotrophomonas Infection | Boolean | Refers to active tissue damage likely invoking some form of inflammatory response caused by Stenotrophomonas | Stenotrophomonas Colonisation (Current), Antibiotic Treatment | Colonisation can lead to infection. Antibiotics can treat and reduce chance of infections |
| Burkholderia Infection | Boolean | Refers to active tissue damage likely invoking some form of inflammatory response caused by Burkholderia | Burkholderia Colonisation (Current), Antibiotic Treatment Regimen | Colonisation can lead to infection. Antibiotics can treat and reduce chance of infections. Burkholderia is, however, harder to treat due to |
| Other Pathogen Infection | Boolean | Refers to active tissue damage likely invoking some form of inflammatory response caused by any other pathogen | Other Pathogen Colonisation (Current), Antibiotic Treatment | Colonisation can lead to infection. Antibiotics can treat and reduce chance of infections |
| Infection | Boolean | Whether there is an active infection of any kind | Diabetes, Pseudomonas Infection, MRSA Infection, Stenotrophomonas Infection, Burkholderia Infection, Other Pathogen | A logical OR of the presence of any infection. Diabetes also increases the overall chance. |
| Inflammation | Boolean | Significant airway or system inflammatory activity | Infection, Anti-inflammatory Treatment Regimen, Fungal Pathogen Colonisation (Current), Noxious Particulates (Current), Abnormal Mucus & Clearance | Infections can give rise to significant inflammation, as can fungal colonisations (without infection). Noxious particulates and abnormal mucus can both directly lead to inflammation (again, without infection). Anti-inflammatories reduce inflammation (chance and severity) |

| Outcomes ||||||
|---|---|---|---|---|
| Variable | Type (Indicative) | Description | Parents | Relationship to Parents |
| Lung Disease Age | Ordered: Child, Adolescent, Young Adult, Adult | The state of disease progression of the lungs and lung function after the exacerbation | Lung Disease Age (Prior), Inflammation, Abnormal Mucus & Clearance | Inflammation and abnormal mucus and clearance can both lead to progression of lung disease age |
| Lung Function | Ordered: Low, Moderate, High | Lung function and capacity after the exacerbation | Lung Disease Age | An increase in the lung disease age likely corresponds to a decline in |
| Cough/Sputum Production | Ordered: Poor, Moderate, Good | A summary of cough and putum production related scores from the CRISS test (e.g., From the questions: How bad was your cough? How much mucus did you cough up?) | Nutrition (BMI), Inflammation, Lung Disease Age (Prior), Lung Disease Age & Function, | Poor nutrition, inflammation, lung disease age and function, and abnormal mucus and clearance can all lead to problematic cough and |
| Dyspnoea | Ordered: Low, Moderate, High | A summary of the dyspnoea related scores from the CRISS test (e.g., How difficult was it to breathe?) | Inflammation, Lung Disease Age (Prior), Lung Disease Age & Function | Inflammation, higher lung disease age and lower lung function can all make shortness of breath more |
| Fatigue | Ordered: Low, Moderate, High | A summary of the fatigue related scores from the CRISS test (e.g., How tired did you feel?) | Inflammation, Diabetes, Lung Disease Age (Prior), Lung Disease Age & | Increases in the degree or chance of any influence make fatigue more likely |
| Mortality | Boolean | Occurrence of death (more importantly, to what extent the risk is elevated after an exacerbation) | Lung Function, Inflammation, Lung Disease Age, Age | A decline in lung function, an increase in lung disease age, older age, or severe inflammation |
| FEV1 | Ordered: Low, Moderate, High | Forced expiratory volume in one second, which is a core measure of lung function | Lung Disease Age & Function | FEV1 is a direct but noisy indicator of lung function |
| CRISS | Ordered: Low, Moderate, High | Cystic Fibrosis Respiratory Symptom Score, a patient-reported metric assessing respiratory | Cough/Sputum Production, Dyspnoea, | CRISS is a noisy summary of all of its parents |



\end{document}